\documentclass[twocolumn,aps,prx,longbibliography]{revtex4-1}
\usepackage{amsmath,latexsym}
\usepackage{xcolor}
\usepackage[%
  colorlinks=true,
  urlcolor=blue,
  linkcolor=blue,
  citecolor=blue
]{hyperref}
\usepackage{etoolbox}
\usepackage{breqn}
\usepackage{mathrsfs,dsfont,verbatim,mathtools,graphicx,amsmath, amssymb, bm, epstopdf,dsfont, enumerate}
\makeatletter
\let\cat@comma@active\@empty
\makeatother

\begin{document}

\title{Intrinsic degree of coherence of classical and quantum states}

\author{Abu Saleh Musa Patoary^{*}, Girish Kulkarni^{*}, and Anand K. Jha}

\email{akjha9@gmail.com}

\affiliation{Department of Physics, Indian Institute of Technology Kanpur, Kanpur, UP 208016, India}
\affiliation{* Both these authors contributed equally}

\date{\today}

\begin{abstract}
In the context of the 2-dimensional (2D) polarization states of light, the degree of polarization $P_{2}$ is equal to the maximum value of the degree of coherence over all possible bases. Therefore, $P_2$ can be referred to as the intrinsic degree of coherence of a 2D state. In addition to (i) the maximum degree of coherence interpretation, $P_2$ also has the following interpretations: (ii) it is the Frobenius distance between the state and the maximally incoherent identity state, (iii) it is the norm of the Bloch-vector representing the state, (iv) it is the distance to the center-of-mass in a configuration of point masses with magnitudes equal to the eigenvalues of the state, (v) it is the visibility in a polarization interference experiment, and (vi) it is the weightage of the pure part of the state. Among these six interpretations of $P_{2}$, the Bloch vector norm, Frobenius distance, and center of mass interpretations have previously been generalized to derive an analogous basis-independent measure $P_{N}$ for $N$-dimensional (ND) states. In this article, by extending the concepts of visibility, degree of coherence, and weightage of pure part to ND spaces, we show that these three remaining interpretations of $P_{2}$ also generalize to the same quantity $P_{N}$, establishing $P_{N}$ as the intrinsic degree of coherence of ND states. We then extend $P_{N}$ to the $N\to\infty$ limit to quantify the intrinsic degree of coherence $P_{\infty}$ of infinite-dimensional states in the orbital angular momentum (OAM), photon number, and position-momentum degrees of freedom. 
\end{abstract}

\maketitle

\section{Introduction} 

Coherence is the physical property responsible for interference phenomena observed in nature and is the subject matter of the classical and quantum theories of coherence \cite{mandel1995cup,born1959pp,zernike1938physica,wolf1959nuovo, glauber1963pr,glauber1963pr2,sudarshan1963prl}. Both these highly-successful theories quantify coherence in terms of the visibility or contrast of the interference. The key difference is that whereas the classical theory formulates the visibility in terms of correlation functions involving products of field amplitudes \cite{born1959pp,wolf1959nuovo, zernike1938physica}, the quantum theory of optical coherence employs correlation functions involving products of field operators that in general may not commute \cite{glauber1963pr,glauber1963pr2,sudarshan1963prl}. In comparison to the classical theory which fails to explain the higher-order correlations of certain quantum light fields \cite{hong1987prl,pittman1996prl}, the quantum theory can be used to quantify the correlations of a general light field to arbitrary orders. However, as far as effects arising from second-order correlations of light fields are concerned, the classical and quantum theories have identical predictions implying that both can be interchangeably used. 

For quantifying the second-order correlations, a quantity of central interest is the degree of coherence, which is just the suitably-normalized second-order correlation function involving electromagnetic fields at two distinct spacetime points or polarization directions \cite{born1959pp,zernike1938physica}. In the context of a partially polarized field represented by a $2\times2$ polarization matrix $\rho$, the degree of coherence is the magnitude of the suitably-normalized off-diagonal entry which quantifies the correlations between the field components along a specific pair of orthogonal polarizations. Thus, the degree of coherence is a manifestly basis-dependent measure of coherence. In contrast, the maximum degree of coherence over all possible orthonormal polarization bases is a basis-independent measure of coherence known as the degree of polarization \cite{wolf1959nuovo}. Owing to this maximum degree of coherence interpretation, we also refer to the degree of polarization $P_{2}$ as the "intrinsic degree of coherence" of the field. For the polarization matrix $\rho$ which is normalized, $P_{2}$ is given by
\begin{align}\label{P2-first}
P_2=\sqrt{2 \ {\rm Tr} \ (\rho^2) -1 }.
\end{align}
In addition to (i) the maximum degree of coherence interpretation, $P_{2}$ also has the following interpretations \cite{born1959pp}: (ii) it is the norm of the Bloch-vector representing the state, (iii) it is the Frobenius distance between the state and the completely incoherent state \cite{luis2005optcomm}, (iv) it is the distance to the center of mass in a configuration of point masses of magnitudes equal to the eigenvalues of the state \cite{alonso2016pra}, (v) it is the visibility obtained in a polarization interference experiment, and (vi) it is the weightage of the completely polarized part of the state. These six interpretations together provide a mathematically appealing and physically intuitive quantification of the intrinsic polarization correlations of a field in a basis-independent manner. 

While the need for a basis-independent quantification of coherence has been recognized long ago in both, the classical and the quantum theories of optical coherence, such a quantification has fully been achieved only for the two-dimensional polarization states of light. In this context, it is known that the $2\times2$ polarization matrix describing the polarization state of a classical light field is formally identical to the $2\times2$ density matrix describing a quantum two-level system. Moreover, there is a one-to-one correspondence between the Poincare sphere representation of partially polarized fields in terms of Stokes parameters \cite{stokes1851trans} and the Bloch sphere representation of qubits in terms of the Bloch vector components \cite{bloch1946pr}. By this correspondence, the measure $P_{2}$ encodes essentially the same information as the quantum purity, and can therefore be used to quantify the intrinsic coherence of both classical and quantum two-dimensional (2D) states \cite{gamel2012pra}. However, a generalized coherence measure analogous to $P_{2}$ that retains all its interpretations has not been obtained for higher-dimensional states so far. 

For quantifying the coherence of higher-dimensional systems, a number of studies in recent years have taken a resource theoretic approach \cite{baumgratz2014prl,girolami2014prl,streltsov2015prl,winter2016prl,streltsov2018njp,ma2019epl}.  However, the present paper does not follow this resource theoretic approach. Instead, it follows an approach from optical coherence theory which seeks to generalize the basis-independent measure of coherence $P_{2}$ and all its known interpretations to quantify the intrinsic degree of coherence of higher-dimensional classical and quantum states.

The first efforts in generalizing $P_{2}$ to higher dimensions were carried out by Barakat \cite{barakat1977optcomm, barakat1983opticaacta} and Samson {\it et al.} \cite{samson1981geophysics, samson1981siam}. In these efforts, they derived a basis-independent measure $P_{N}$ for an $N\times N$ polarization matrix $\rho$ by generalizing the Bloch-vector norm interpretation of $P_2$ to an ND space. In particular, they showed that for a normalized $\rho$,
\begin{align}
P_N=\sqrt{\frac{N\,{\rm Tr}(\rho^2)-1}{N-1}} \label{P_N}.
\end{align}
Recently, following up on previous generalizations for three- \cite{luis2005pra} and four- \cite{luis2007josaa} dimensional spaces, the Frobenius distance interpretation of $P_2$ \cite{luis2005optcomm} was generalized to ND spaces to also yield $P_{N}$ \cite{yao2016scirep}. In addition, the center of mass interpretation when applied to ND states yields $P_{N}$ as the generalized measure. Thus, it has so far been possible to show that $P_{N}$ has three of the six interpretations of $P_2$. However, the generalization of the remaining three interpretations have either not been attempted or have had limited success \cite{ellis2005optcomm, gamel2012pra, setala2009optlett}. In this article, we take up the other three interpretations of $P_2$, namely, the visibility, degree of coherence, and weightage of pure part interpretations and extend them to ND spaces. We show that even these three interpretations of $P_{2}$ generalize to the same measure $P_{N}$. In essence, by demonstrating that $P_{N}$ has all the six interpretations of $P_{2}$, we theoretically establish $P_{N}$ as quantifying the intrinsic degree of coherence of ND states. We then extend $P_{N}$ to the $N\to\infty$ limit to quantify the intrinsic degree of coherence $P_{\infty}$ of infinite-dimensional states.  

The paper is organized as follows. In Sec.~II, we present a conceptual description of the degree of polarization. In Sec.~III, we describe the existing work on how the expression for $P_N$ is obtained by generalizing the Bloch-vector norm, Frobenius distance, and center of mass interpretations of $P_2$ to $N$-dimensional states. In Sec.~IV, we generalize the concepts of visibility, degree of coherence, and weightage of pure part to ND spaces, demonstrate that each of these interpretations of $P_{2}$ uniquely generalizes to $P_{N}$, and thereby establish $P_N$ as the  intrinsic degree of coherence of finite $N$-dimensional classical and quantum states. In Sec.~V, we consider infinite-dimensional states in the orbital angular momentum (OAM), photon number, position and momentum bases, and show that the intrinsic degree of coherence $P_{\infty}$ of a normalizable state $\rho$ is given by $P_{\infty}=\sqrt{\mathrm{Tr}(\rho^2)}$. In the rest of the paper, we will use the symbol $\rho$ to denote the density matrix of dimensionality $2, N$ or $\infty$ depending on the context. Also, we will denote the $N\times N$ identity matrix by $\mathds{1}_{N}$.

\section{Degree of Polarization}

The polarization state of an electromagnetic field can be represented by a positive-semidefinite $2\times2$ Hermitian matrix. It is referred to as the polarization matrix or the coherence matrix and is defined as \cite{mandel1995cup},
\begin{align}
 \rho=\begin{bmatrix}
\langle E_{1}E^*_{1}\rangle & \langle E_{1}E^*_{2}\rangle \\
\langle E^*_{1}E_{2}\rangle & \langle E_{2}E^*_{2}\rangle
\end{bmatrix}=\begin{bmatrix}
\rho_{11} & \rho_{12} \\
\rho_{21} & \rho_{22}
\end{bmatrix}. \label{polarization matrix}
\end{align}
Here $\langle\cdots\rangle$ denotes the ensemble average over many realizations of the field, and $E_{1}$ and $E_{2}$ denote the electric field components along two mutually orthonormal polarization directions represented by the basis vectors $\{|1\rangle$, $|2\rangle\}$, and $\rho_{ij}$ with $i,j=1,2$ denote the matrix elements of $\rho$ in the $\{|1\rangle,|2\rangle \} $ basis. The basis-dependent quantity $\mu_2=|\rho_{12}|/\sqrt{\rho_{11}\rho_{22}}$ is called the degree of coherence between the polarization basis vectors $\{|1\rangle$, and $|2\rangle \}$. It was shown by Wolf in a classic paper that the maximum value of $\mu_2$ over all possible choices of the bases in the 2D Hilbert space is equal to the degree of polarization $P_2$ \cite{wolf1959nuovo}, which for a normalized $\rho$ can be shown to be \cite{born1959pp}:
\begin{align}\label{P2-wolf}
P_2=\sqrt{1-4\,{\rm det}\,\rho}=\sqrt{2 \ {\rm Tr} \ (\rho^2) -1 }.
\end{align}
As the trace and the determinant are invariant under unitary operations, $P_2$ is a basis-independent quantity. Furthermore, $0\leq P_{2}\leq1$ with $P_{2}=1$ only when $\rho$ is a perfectly polarized field (pure state) and $P_{2}=0$ only when $\rho$ is the completely unpolarized field (completely mixed state) represented by the identity matrix. In the next two sections, we consider the six known interpretations of $P_{2}$ that justify its suitability as an intrinsic degree of coherence for 2D states. Following a brief description of each interpretation, we present the generalization to ND space and obtain $P_{N}$ as the ND analog of $P_{2}$.

\section{Existing works on generalizing interpretations of $P_{2}$ to ND states}

\subsection{Bloch vector norm interpretation}
 
\subsubsection*{\bf 2D states} It is known that an arbitrary 2D state $\rho$ has the following unique decomposition in terms of the Stokes parameters \cite{nielsenchuang}:
\begin{align}\label{2D-stokes-decomp}
 \rho=\frac{1}{2}\Big(\mathds{1}_{2}+\sum_{i=1}^{3} r_{i}\sigma_{i}\Big).
\end{align}
Here $\sigma_{1},\sigma_{2}$ and $\sigma_{3}$ are the Pauli matrices, and the real scalar quantities $r_{i}$'s are called the Stokes parameters of the state. Such a parametrization is possible due to the fact that $\sigma_{i}$'s, which are the generators of the Lie group $\rm SU(2)$, form an orthonormal basis in the real vector space of traceless $2\times2$ Hermitian matrices with respect to the Hilbert-Schmidt inner-product, $(A,B)\equiv \mathrm{Tr}\left(A^{\dagger}B\right)$. Consequently, the parameters $r_{i}$ can be regarded as the components of a $3$-dimensional vector $\bm{r}\equiv(r_{1},r_{2},r_{3})$, which is referred to as the Bloch vector representing the state in this vector space. For a 2D density matrix $\rho$, the condition $\mathrm{Tr}\,\rho^2\leq1$ is both necessary and sufficient to ensure positive-semidefiniteness, which in turn implies that the space of physical states is characterized by $0\leq|\bm{r}|\leq 1$. This space can be imagined to be a closed sphere in 3 dimensions, termed as the Bloch sphere. The pure states reside on the surface of this sphere with $|\bm{r}|=1$, whereas the maximally incoherent state $\mathds{1}_{2}/2$ with $|\bm{r}|=0$ resides at the center. From Eq.~(\ref{P2-wolf}), it can be shown that the norm of the Bloch vector is equal to $P_{2}$, that is, $|\bm{r}|=\sqrt{\sum_{i=1}^{3}|r_{i}|^2}=P_{2}$ \cite{nielsenchuang}. This way, $P_{2}$ is interpreted as the norm of the Bloch vector representing the state.

\subsubsection*{\bf ND states} 

In direct correspondence with Eq.~(\ref{2D-stokes-decomp}), it has been shown that any ND state $\rho$ can be decomposed as \cite{hioe1981prl,kimura2003pla,byrd2003pra,bertlmann2008jpa},
\begin{align}\label{Ndim-Bloch-form}
 \rho=\frac{1}{N}\Big(\mathds{1}_{N}+\sqrt{\frac{N(N-1)}{2}}\sum_{i=1}^{(N^2-1)} r_{i}\Lambda_{i}\Big),
\end{align}
where $\Lambda_{i}$'s are the generalized $N\times N$ Gellmann matrices, and the scalar quantities $r_{i}$'s are the ND analogs of Stokes parameters. In exact analogy with the 2D case, this parametrization is made possible by the fact that $\Lambda_{i}$'s, which are the $(N^2-1)$ generators of the Lie group $\rm SU(N)$, form an orthonormal basis in the real vector space of traceless $N\times N$ Hermitian matrices with respect to the Hilbert-Schmidt inner-product. The parameters $r_{i}$ form the components of the $(N^2-1)$-dimensional Bloch vector $\bm{r}$ representing the state $\rho$. We note that in contrast with the 2D case, the condition $\mathrm{Tr}\,\rho^2\leq1$ is not sufficient to ensure positive-semidefiniteness of ND density matrices. Consequently, only a subset of states represented by the $(N^2-1)$-dimensional sphere and defined by $0\leq|\bm{r}|\leq1$ correspond to physical states \cite{kimura2003pla,byrd2003pra}.

Barakat \cite{barakat1977optcomm, barakat1983opticaacta} and Samson {\it et al.} \cite{samson1981geophysics, samson1981siam} were the first ones to show that the norm of the ND Bloch-vector is the degree of polarization $P_N$ of the state. The derivations of $P_N$ by both Barakat \cite{barakat1977optcomm, barakat1983opticaacta} and Samson {\it et al.} \cite{samson1981geophysics, samson1981siam} were presented in terms of the eigenvalues of $\rho$ and not in terms of the Gellman matrices. For 3D states, an explicit derivation of $P_3$ in terms of 3D Gellman matrices was carried out by Set\"{a}l\"{a} \textit{et al.} \cite{setala2002prl, setala2002pre} who also demonstrated usefulness of $P_3$ for studying optical near fields and evanescent fields. We now present the derivation for ND states explicitly in term of ND Gellman matrices and obtain the expression of $P_{N}$ as in Eq.~(\ref{P_N}).

We note that the set of $(N^2-1)$ generalized Gellmann matrices $\Lambda_{i}$'s of Eq.~(\ref{Ndim-Bloch-form}) comprises of  three subsets: the set $\{U\}$ of $N(N-1)/2$ symmetric matrices, the set $\{V\}$ of $N(N-1)/2$ anti-symmetric matrices, and the set $\{W\}$ of $(N-1)$ diagonal matrices. The explicit forms of these matrices in the orthonormal basis $\{|i\rangle\}^{N}_{i=1}$, where $|i\rangle$ is an ND column vector with the $i^{\rm th}$ entry being 1 and others being 0, are given by \cite{kimura2003pla}, 
\begin{align}
&U_{jk}=|j\rangle \langle k|+|k\rangle \langle j|, \qquad V_{jk}=-i|j\rangle \langle k|+i|k\rangle \langle j|, \notag \\
& {\rm and} \ \ W_{l}=\sqrt{\frac{2}{l(l+1)}}\Big(\sum_{m=1}^l|m\rangle \langle m|-l|l+1\rangle \langle l+1|\Big).\label{Generalized Gellmann}
\end{align}
where $1\leq j<k\leq N$ and $1\leq l\leq (N-1)$. In terms of these definitions, we write Eq.~(\ref{Ndim-Bloch-form}) as,
\begin{multline}\label{Ndim-Bloch-form-2}
 \rho=\frac{1}{N}\Bigg[\mathds{1}_{N}+\sqrt{\frac{N(N-1)}{2}}\Big(\sum_{j=1}^N\sum_{k=j+1}^N \big\{u_{jk}U_{jk}\big.\\\big.+v_{jk}V_{jk}\big\}+\sum_{l=1}^{N-1}w_{l}W_{l}\Big)\Bigg],
\end{multline}
where $u_{jk},v_{jk}$ and $w_{l}$ are the Bloch-vector components along the Gellmann matrices $U_{jk},V_{jk}$ and $W_{l}$ respectively. Here, we have relabeled the set of components $\{r_{i}\}$ and the set of matrices $\{\Lambda_{i}\}$ of Eq.~(\ref{Ndim-Bloch-form}) by the set of parameters $\{\{u_{jk}\},\{v_{jk}\},\{w_{l}\}\}$ and the set of matrices $\{\{U_{jk}\},\{V_{jk}\},\{W_{l}\}\}$, respectively. We calculate the components $u_{jk},v_{jk}$ and $w_{l}$ in terms of the density matrix elements and find them to be
\begin{align}
&u_{jk}=\sqrt{\frac{N}{2(N-1)}}(\rho_{jk}+\rho_{kj}), \notag \hspace{4mm} v_{jk}=i\sqrt{\frac{N}{2(N-1)}}(\rho_{jk}-\rho_{kj}), \notag \\
&w_{l}=\sqrt{\frac{N}{l(l+1)(N-1)}}\Big(\sum_{m=1}^l\rho_{mm}-l\rho_{l+1,l+1}\Big). \label{coeff    s}
\end{align}
The norm of the Bloch vector $\bm{r}$ defined as $|\bm{r}|=\sqrt{\sum_{i=1}^{(N^2-1)}r^2_{i}}$ is therefore given by, 
\begin{align}\label{bloch-norm}
 |\bm{r}|=\sqrt{\sum_{j=1}^N\sum_{k=j+1}^N\Big[u^2_{jk}+v^2_{jk}\Big]+\sum_{l=1}^{N-1}w^2_l}.
\end{align}
In order to evaluate $|\bm{r}|$, we first find that 
\begin{align}\label{coeff uv}
\sum_{j=1}^N\sum_{k=j+1}^N\Big[u^2_{jk}+v^2_{jk}\Big] = \frac{2N}{N-1}\sum_{j=1}^N\sum_{k=j+1}^N |\rho_{jk}|^2.
\end{align}
We then evaluate the other summation in Eq.~(\ref{bloch-norm}) to be
\begin{align} \label{coeff w} \nonumber
&\sum_{l=1}^{N-1}w^2_l=\sum_{l=1}^{N-1}\frac{N}{l(l+1)(N-1)}\Big(\sum_{m=1}^l\rho_{mm}-l\rho_{l+1,l+1}\Big)^2 \\ \nonumber
  & \!\begin{multlined}= \frac{N}{N-1}\Big[\sum_{i=1}^N\rho_{ii}^2\big\{\sum_{j=i}^{N-1}\frac{1}{j(j+1)}+\frac{i-1}{i}\big\}-\frac{2}{N}\sum_{i=1}^N\sum_{j=i+1}^N\rho_{ii}\rho_{jj}\Big]
 \end{multlined}\\
& =\sum_{i=1}^N\rho_{ii}^2-\frac{2}{N-1}\sum_{i=1}^N\sum_{j=i+1}^N\rho_{ii}\rho_{jj}.
\end{align}
By substituting Eqs.~(\ref{coeff uv}) and (\ref{coeff w}) into Eq.~(\ref{bloch-norm}), we obtain
\begin{align}
 |\bm{r}|=P_N=\sqrt{\frac{N\,{\rm Tr}(\rho^2)-1}{N-1}}=P_{N}.
\end{align}
Thus $P_N$, like its two-dimensional analog, can be interpreted as the norm of the Bloch vector corresponding to the ND state.

\subsection{Frobenius distance interpretation}

\subsubsection*{\bf 2D states} 

For a 2D state $\rho$, it was known that the degree of polarization $P_2$ can be viewed as the Frobenius-distance between the state $\rho$ and the completely-incoherent state $\mathds{1}_{2}/2$  \cite{luis2005optcomm}, that is,
\begin{align}\label{P2-as-frob-norm}
P_2=\sqrt{2}\Big{|}\Big{|}\rho-\frac{\mathds{1}_{2}}{2}\Big{|}\Big{|}_F = \sqrt{2 \ {\rm Tr} \ (\rho^2) -1 }.
\end{align} 
Here, the Frobenius-distance is quantified using the Frobenius-norm, defined as $||A||_F\equiv \sqrt{{\rm Tr}(A^{\dagger}A)}$, with the normalization factor ensuring that $0\leq P_{2}\leq1$. We see that the expressions of $P_2$ in Eq.(\ref{P2-wolf}) and Eq.(\ref{P2-as-frob-norm}) are same.

\subsubsection*{\bf ND states}

The Frobenius-distance interpretation was first generalized to three- \cite{luis2005pra} and four- \cite{luis2007josaa} dimensional states by A.~Luis. More recently,  Yao \textit{et al.} \cite{yao2016scirep} have generalized the Frobenius-distance interpretation to ND states to define $P_N$ as:
\begin{align}\label{PN-as-frob-norm}
P_N\equiv\sqrt{\frac{N}{N-1}}\Big{|}\Big{|}\rho-\frac{\mathds{1}_{N}}{N}\Big{|}\Big{|}_F = P_N=\sqrt{\frac{N\,{\rm Tr}(\rho^2)-1}{N-1}}.
\end{align}
In other words, $P_{N}$ is the Frobenius-distance between the state $\rho$ and the completely-incoherent state $\mathds{1}_{N}/N$ in the space of $N\times N$ density matrices. The normalization factor in Eq.~(\ref{PN-as-frob-norm}) is again chosen such that $0\leqslant P_N \leqslant 1$. We note that the expressions of $P_N$ in Eqs.~(\ref{P_N}) and (\ref{PN-as-frob-norm}) are the same. Furthermore, it can be verified that when $\rho$ is pure, $\mathrm{Tr}(\rho^2)=1$ implying $P_{N}=1$, whereas when $\rho=\mathds{1}_{N}/{N}$, $\mathrm{Tr}(\rho^2)=1/N$ implying $P_{N}=0$. 

\subsection{Center of mass interpretation}
In a recent study, M. A. Alonso \textit{et. al} \cite{alonso2016pra} have discussed a geometric interpretation of the measure $P_{N}$ of Eq.~(\ref{P_N}) as the distance to the center of mass in a configuration of point masses.
\subsubsection*{\bf 2D states} 
 Consider a configuration of 2 point masses of magnitudes equal to the eigenvalues $\lambda_{1}$ and $\lambda_{2}$ of the state, each placed at a unit distance from the origin in opposite directions in a 1-dimensional Euclidean space. The distance $Q$ to the center of mass of this configuration from the origin is given by
\begin{equation}
 Q=\Big|\frac{\lambda_{1}-\lambda_{2}}{\lambda_{1}+\lambda_{2}}\Big|=P_{2}.
\end{equation}
Thus, $P_{2}$ has the interpretation as the distance of the center of mass from the origin in this configuration.
\subsubsection*{\bf ND states} 
Consider a configuration of $N$ point masses of magnitudes equal to the eigenvalues $\lambda_{1},\lambda_{2},...,\lambda_{N}$ of $\rho$, each placed at a unit distance from the origin and equally-spaced from one-another such that they constitute a regular $(N-1)$-simplex in an $(N-1)$-dimensional Euclidean space. The distance $Q$ to the center of mass of this configuration is given by
\begin{equation}
 Q=\sqrt{\frac{\sum_{i=1}^{N-1}\sum_{j=i+1}^N(\lambda_i-\lambda_j)^2}{(N-1)(\sum_{i=1}^N\lambda_i)^2}}=P_{N}.
\end{equation}
Therefore, $P_{N}$ is equal to the distance of the centre of mass of this configuration from the origin. 
\section{Generalizing other interpretations of $P_2$ to ND states}

\subsection{Maximum degree of coherence interpretation}

\subsubsection*{\bf 2D states} 

As pointed out in Sec.~II in the context of 2D polarization sates,  the basis-dependent quantity $\mu_2$ in Eq.~(\ref{polarization matrix}) quantifies the degree of coherence between the mutually orthogonal polarizations states represented by $|1\rangle$ and $|2\rangle$. Using Eqs.~(\ref{P2-wolf}) and (\ref{mu2}), it can be shown that $0\leq\mu_{2}\leq P_{2}$ and also that $\mu_{2}$ attains the maximum value $P_2$ when the basis  $\{|1\rangle, |2\rangle\} $ is such that $\rho_{11}=\rho_{22}$ \cite{born1959pp,wolf1959nuovo}, that is,
\begin{align}\label{mu2-max}
\max_{\{|1\rangle, |2\rangle\}\in \mathds{S}} \mu_{2}=P_{2}.
\end{align}
In this way, $P_{2}$ is interpreted as the maximum of $\mu_{2}$ over the set $\mathds{S}$ of all orthonormal bases in the $2$D Hilbert space. In order to generalize the definition of the degree of coherence for ND states, we rewrite $\mu_2$ as
\begin{align} \label{mu2}
\mu_{2}=\sqrt{\frac{|\rho_{12}|^2}{\rho_{11}\rho_{22}}}.
\end{align}
We find that while the numerator $|\rho_{12}|^2$ quantifies the correlation between the basis vectors $|1\rangle$ and $|2\rangle $, the denominator provides the normalization such that $0\leq\mu_2\leq 1$. Our aim is to define an ND degree of coherence $\mu_N$ such that it reduces to $\mu_2$ for $N=2$ and  lies between 0 and 1.

\subsubsection*{\bf ND states} 

We use the definition in Eq.~(\ref{mu2}) to generalize the concept of the degree of coherence to ND states. We expect the generalized quantity $\mu_N$ to be basis-dependent, the maximum of which must be equal to the ND intrinsic degree of coherence $P_N$. Therefore, in analogy with the definition of $\mu_2$ in Eq.~(\ref{mu2}), we define the ND degree of coherence $\mu_N$ as
\begin{align} \label{muN}
\mu_N = \sqrt{\frac{\sum_{i=1}^{N-1} \sum_{j=i+1}^N|\rho_{ij}|^2}{\sum_{i=1}^{N-1}\sum_{j=i+1}^N \rho_{ii}\rho_{jj}}}.
\end{align}
Here, $\rho_{ij}$ are the matrix elements of the state $\rho$ in an orthonormal basis $\{|1\rangle,|2\rangle,\cdots,|N\rangle\}$. The numerator is the sum of the squared magnitudes of all the off-diagonal terms and the denominator is the sum of the products of the pairs of diagonal terms. As expected, $\mu_N$ as defined above reduces to $\mu_2$ for $N=2$, and the normalization term in the denominator makes sure that $\mu_N$ lies between 0 and 1. We further note that $\mu_N$ is a basis-dependent quantity. Now, in order for $\mu_N$ to be considered as the ND analog of $\mu_{2}$, we need to show that the maximum value of $\mu_N$ over the set of all possible ND bases is equal to $P_N$.  From Eq.~(\ref{PN-as-frob-norm}) and Eq.~(\ref{muN}), we have
\begin{align} \label{muN2}
\mu_N^2 &= \frac{\sum_{i=1}^{N-1} \sum_{j=i+1}^N|\rho_{ij}|^2}{\sum_{i=1}^{N-1}\sum_{j=i+1}^N \rho_{ii}\rho_{jj}}=\frac{\frac{1}{2}\left(\sum_{i=1}^{N} \sum_{j=1}^N|\rho_{ij}|^2 -\sum_{i=1}^N \rho_{ii}^2\right)}{\frac{1}{2}\left(\sum_{i=1}^{N}\sum_{j=1}^N \rho_{ii}\rho_{jj} - \sum_{i=1}^N \rho_{ii}^2 \right)}  \notag \\
&= \frac{{\rm Tr}(\rho^2)-\sum_{i=1}^N \rho_{ii}^2}{1 - \sum_{i=1}^N \rho_{ii}^2}=1-\frac{1-{\rm Tr}(\rho^2)}{1-\sum_{i=1}^N \rho_{ii}^2}.
\end{align}
From the above equation, it is clear  that $\mu_N^2$ attains its minimum value when the sum  $\sum_{i=1}^N \rho_{ii}^2$ is maximum. The sum is maximum when $\rho_{ii}$ is equal to 1 only for a particular $i$ and is zero for the rest, in which case the sum $\sum_{i=1}^N \rho_{ii}^2={\rm Tr}(\rho^2)$ implying ${\rm min} \ \mu_N=0$. Furthermore, $\mu_N^2$ attains its maximum value when the sum  $\sum_{i=1}^N \rho_{ii}^2$ is minimum. It is straightforward to show that the sum $\sum_{i=1}^N \rho_{ii}^2$ is minimum when  $\rho_{11}=\rho_{22}=\cdots=\rho_{NN}=1/N$, in which case $\sum_{i=1}^N \rho_{ii}^2=\sum_{i=1}^N (1/N)^2=1/N$. Therefore, from Eq.~(\ref{muN2}), we have 
\begin{align}\label{muN-max}
\max_{\{|1\rangle, |2\rangle, \cdots |N\rangle \}\in \mathds{S}} \ \mu_N &=\sqrt{\frac{N \ {\rm Tr}(\rho^2)-1}{N-1}}=P_N,
\end{align}
which is in direct correspondence with Eq.~(\ref{mu2-max}). Thus, as in the 2D case, we find that the maximum of $\mu_{N}$ over the set $\mathds{S}$ of all orthonormal bases in the ND Hilbert space is equal to the intrinsic degree of coherence $P_{N}$. Moreover, the maximum is achieved in the basis where all the diagonal entries are equal, again as is true in the 2D case. While our analysis does not present a clear physical reasoning for defining $\mu_{N}$ as Eq.~(\ref{muN}), the fact that $\mu_{N}$ satisfies all the mathematical properties of $\mu_{2}$ strongly suggests that $\mu_{N}$ is the ND analog of $\mu_{2}$, and can therefore be referred to as the ND degree of coherence.

We now note that our above analysis is physically distinct from a recent study \cite{streltsov2018njp} which relates the maximal resource-theoretic coherence of a state over unitary transformations to the state purity. The distinction arises because whereas optical coherence theory quantifies the system's ability to interfere, the resource theory of coherence quantifies the amount of superposition in a specific basis that can be exploited for certain quantum protocols. In order to illustrate this difference in the context of a 2D state $\rho$, we consider the $l_{1}$-norm measure $|\rho_{12}|$ from resource theory, and the degree of coherence $\mu_{2}=|\rho_{12}|/\sqrt{\rho_{11}\rho_{22}}$ of Eq.~(\ref{mu2}) from optical coherence theory. For a pure state $\rho=|\psi\rangle\langle\psi|$, where $|\psi\rangle=\epsilon |1\rangle+\sqrt{1-\epsilon^2}|2\rangle$ with $\epsilon\to0$, we have $|\rho_{12}|\to0$ which implies that the state is incoherent in a resource-theoretic sense, whereas $\mu_{2}=1$ which implies that the state is fully coherent in the optical coherence-theoretic sense. Therefore, while it is interesting that similar relations between maximal coherence and purity hold in both theories, these relations are physically distinct.

\subsection{Visibility Interpretation}

\subsubsection*{\bf 2D states} 

\begin{figure}[t!]
\centering
\includegraphics[width=85mm]{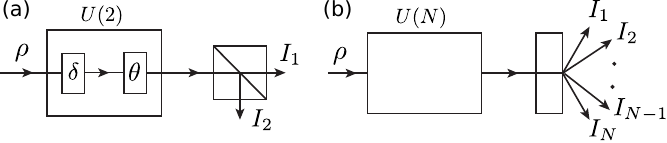}
\caption{(a) Schematic setup for describing degree of polarization $P_2$ as the visibility in a polarization interference experiment. (b) Schematic setup for describing $N$-dimensional degree of polarization or $N$-dimensional intrinsic degree of coherence $P_{N}$ as the $N$-dimensional visibility in an interference experiment. PBS stands for polarizing beam splitter and NPS stands for $N$-port splitter.}\label{fig1}
\end{figure}

The visibility interpretation of $P_2$ for a 2D state was given by Emil Wolf \cite{wolf1959nuovo} using a polarization interference scheme (see Section 6.2 of Ref.~\cite{mandel1995cup}). As depicted in Fig.~\ref{fig1}, we discuss this scheme with slight modifications in order to make it more amenable to generalization to higher dimensions. A field in the polarization state $\rho$, as given by Eq.~(\ref{polarization matrix}), first passes through a wave-plate (WP) that introduces a phase $\delta$ between the two mutually orthogonal directions represented by vectors $|1\rangle$ and $|2\rangle$. The field then passes through a rotation plate (RP) that rotates the polarization state by an angle $\theta$. 
%
%
%
Finally, the field is detected using the polarizing beam splitter (PBS) in the two orthogonal polarization directions $|1\rangle$ and $|2\rangle$. The corresponding detection probabilities $I_1$ and $I_2$ at the two output ports are given by
\begin{align}
&I_1=\rho_{11}\cos^2\theta+\rho_{22}\sin^2\theta+|\rho_{12}|\sin\theta\cos\theta\cos(\beta+\delta),\notag \\
&I_2=\rho_{11}\sin^2\theta+\rho_{22}\cos^2\theta-|\rho_{12}|\sin\theta\cos\theta\cos(\beta+\delta) \notag,
\end{align}
where $\rho_{12}=|\rho_{12}| e^{i\beta}$. The visibility $V$ of the interference pattern is defined as (see Section 6.2 of Ref.~\cite{mandel1995cup}) 
\begin{align}\label{visibility-wolf}
V=\frac{\langle I_1 \rangle_{{\rm max}(\delta, \theta)}-\langle I_1 \rangle_{{\rm min}(\delta, \theta)}}{\langle I_1 \rangle_{{\rm max}(\delta, \theta)}+\langle I_1 \rangle_{{\rm min}(\delta, \theta)}},
\end{align}
where $\langle I_1 \rangle_{{\rm max}(\delta, \theta)}$ and $\langle I_1 \rangle_{{\rm min}(\delta, \theta)}$ are the maximum and minimum values of $I_{1}$, respectively, over all possible $\delta$ and $\theta$. Similarly, we can equivalently define the visibility as
\begin{equation}\label{visibility}
V=\max_{U\in U(2)}\Big|\frac{I_1-I_2}{I_1+I_2}\Big|=\max_{U\in U(2)}f(I_1,I_2),
\end{equation}
where $U(2)$ is the group of 2D unitary matrices and where we have denoted $|(I_1-I_2)/(I_1+I_2)|$ as $f(I_1, I_2)$ since we would find this notation to be more convenient when generalizing to ND spaces. The function $f(I_1, I_2)$ has the following properties: (i) It is $1$ if and only if one among $I_1$ and $I_2$ is 1 and the other one is 0, (ii) It is 0 if and only if $I_1=I_2$, (iii) It is a {\it Schur-convex} function, that is, for two given sets of probabilities $\{I_{1},I_2\}$ and $\{I'_{1},I'_2\}$ if $\{I'_1,I'_2\}$ majorizes $\{I_1,I_2\}$ then $f(I_1,I_2)\leq f(I'_1,I'_2)$ \cite{bhatia2013springer}. The maximization involved in Eq.~(\ref{visibility}) can be carried out using Schur's theorem which states that the measured probability distribution of a state in any basis is majorized by the eigenvalue distribution of the state \cite{nielsen2002notes}, that is, $(I_{1},I_{2})\prec(\lambda_{1},\lambda_{2})$. Since there always exists a unitary transformation such that $I_{1}=\lambda_{1}$ and $I_{2}=\lambda_{2}$, $f(I_1, I_2)$ becomes maximum when $I_{1}=\lambda_{1}$ and $I_{2}=\lambda_{2}$, and in that case we get  
\begin{equation}\label{eigen-visibility}
V=\max_{U\in U(2)}f(I_1,I_2)=f(\lambda_1,\lambda_2)=\Big|\frac{\lambda_{1}-\lambda_{2}}{\lambda_{1}+\lambda_{2}}\Big|=P_{2},
\end{equation}
that is, $P_2$ equals the 2D visibility in a polarization interference experiment. The importance of the visibility interpretation is that it not only provides a physically intuitive way of understanding the degree of coherence but also provides an experimental scheme for measuring it.

\subsubsection*{\bf ND states} 

In direct analogy with the scheme depicted in Fig.~\ref{fig1}(a), Fig.~\ref{fig1}(b) depicts the general interference situation for an ND density matrix $\rho$ represented in an orthonormal basis $\{|1\rangle,|2\rangle,\cdots,|N\rangle\}$. The density matrix $\rho$ is acted upon by a general $N \times N$ unitary operator $U$, which can be realized by a combinations of optical elements. The $N$-port splitter (NPS) divides the density matrix along $N$ orthonormal states $\{|1\rangle,|2\rangle,\cdots,|N\rangle\}$ and the the detection probabilities along the basis vectors are represented by  $\{I_{1},I_{2},\cdots,I_{N}\}$. In analogy with the definition of $f(I_{1},I_{2})$ for the 2D case, we define
\begin{align} \label{ND Visibility}
f(I_1,I_2,\cdots,I_N)=\sqrt{\frac{\sum_{i=1}^{N-1}\sum_{j=i+1}^N(I_i-I_j)^2}{(N-1)(\sum_{i=1}^N I_i)^2}},
\end{align}
which satisfies the following properties: (i) It is $1$ if and only if $I_{i}=1$ for some $i=k$, and $I_{i}=0$ for $i\neq k$, where $i=1, 2, \cdots, N$ and $k\leq N$, (ii) It is $0$ if and only if all the probabilities are equal, that is, $I_{i}=1/N,\,\,\,{\rm where}\ i=1,2,\cdots,N$ and (iii) It is a {\it Schur-convex} function, as may be proved using theorem II.3.14 of Ref.~\cite{bhatia2013springer}. We know by virtue of Schur's theorem \cite{nielsen2002notes} that $\{I_{1},I_{2},\cdots,I_{N}\}\prec\{\lambda_{1},\lambda_{2},\cdots,\lambda_{N}\}$, where $\lambda_i$s are eigenvalues of the density matrix. We also know that there always exists a unitary transformation $U\in U(N)$ such that $\{I_{1},I_{2},\cdots,I_{N}\}=\{\lambda_{1},\lambda_{2},\cdots,\lambda_{N}\}$.  Using these facts, we define the ND visibility $V$ as $f(I_1,I_2\cdots,I_N)$ maximized over $U(N)$, i.e, 
\begin{align}
V &=\max_{U\in U(N)} f(I_1,I_2\cdots,I_N)=\sqrt{\frac{\sum_{i=1}^{N-1}\sum_{j=i+1}^N(\lambda_i-\lambda_j)^2}{(N-1)(\sum_{i=1}^N\lambda_i)^2}} \notag \\
 &=\sqrt{\frac{N \ {\rm Tr}(\rho^2)-1}{N-1}}=P_{N}. \label{PN-visibility}
\end{align}
Thus, we find that just as in the 2D case, $P_N$ has the interpretation as the ND visibility of an experiment. 

\subsection{Weightage of pure part interpretation}

\subsubsection*{\bf 2D states} 

In the context of partially polarized fields, it has been shown that any 2D polarization state $\rho$ can be uniquely decomposed 
into a weighted mixture of two fields, one of which is completely polarized or pure, and the other one completely unpolarized or fully mixed \cite{mandel1995cup, born1959pp}. Mathematically , this implies that
%
%
%
%
%
\begin{align}\label{wolf-decomp}
\rho=s_{1} |\psi_{1}\rangle\langle\psi_{1}| + (1-s_{1})\frac{\mathds{1}_{2}}{2},
\end{align}
where $|\psi_{1}\rangle$ represents the completely polarized pure state, $s_1=\lambda_1-\lambda_2$ with $\lambda_{1}$ and $\lambda_{2}$ being the eigenvalues of $\rho$ denotes the weightage of the pure part, and $\mathds{1}_{2}$ is the completely unpolarized state. From Eq.~(\ref{eigen-visibility}), we know that for a normalized $\rho$, $\lambda_1 -\lambda_2=P_2$, from which we get 
\begin{align}
s_{1}=\lambda_{1}-\lambda_{2}=P_{2}.
\end{align}
In other words, $P_2$ is equal to the weightage of the pure portion of the state. This interpretation is physically intuitive as it implies that in order to prepare the state by mixing together a pure state and the completely mixed state, the needed weightage of the pure part is $P_{2}$.

\subsubsection*{\bf ND states}

We now generalize this interpretation of $P_{2}$ to higher dimensions. The quantification of $P_2$ in terms of the weightage of its pure part is possible only because of the existence of the unique decomposition in Eq.~(\ref{wolf-decomp}). However, it is now known that such a unique decomposition in terms of just two matrices is not possible for ND states  \cite{brosseau1998fundamentals, gill2014pra,gill2017pra}. For a 3D polarization state it has been shown that a unique decomposition is possible in terms of three matrices, one of which is the rank-1 matrix which is a pure state, the second one is a rank-2 matrix and the third one is the identity matrix \cite{ellis2005optcomm}. It has been argued that the weightage of the pure part of this decomposition, which is equal to $\lambda_1-\lambda_2$, where $\lambda_1$ and $\lambda_2$ are the two largest eigenvalues of $\rho$, could be taken as the degree of polarization of the 3D state. However, a few issues have been pointed out regarding this decomposition because of which the weightage of the rank-1 matrix of this decomposition cannot in general be taken as the 3D degree of polarization \cite{gamel2012pra, setala2009optlett}.

In contrast, we now show that it is possible to have a unique decomposition of an ND state as a weighted mixture of $N$ matrices as given below, one of which is completely mixed and the rest $N-1$ are completely pure.
\begin{equation}\label{PN-decomp}
 \rho=\sum_{i=1}^{N-1}s_{i}|\psi_{i}\rangle\langle\psi_{i}|+\Big(1-\sum^{N-1}_{i=1}s_{i}\Big)\frac{\mathds{1}_{N}}{N},
\end{equation} 
Here the states $\{|\psi_{i}\rangle\}$'s are pure and orthonormal and the corresponding weightages $s_{i}$'s are real and non-negative. In order to ensure a unique decomposition for every physical density matrix, it must be verified that the number of independent parameters are identical on the two sides of Eq.~(\ref{PN-decomp}). On the left side, the density matrix $\rho$ has $(N^2-1)$ free parameters. On the right side:(i) there are $(N-1)\,\,s_{i}$'s, (ii) each of the $(N-1)\,\,|\psi_{i}\rangle$'s has $2(N-1)$ free parameters, and (iii) the mutual orthogonality between $|\psi_{i}\rangle$'s would introduce $(N-1)(N-2)$ constraints. These conditions imply $(N^2-1)$ free parameters on the right-hand side as well. We introduce an additional vector $|\psi_{N}\rangle$ to the set of $(N-1)\,\,|\psi_{i}\rangle$'s such that $|\psi_i\rangle$ with $i=1\cdots N$ form an orthonormal and complete basis, that is, $\sum_{i=1}^{N}|\psi_i\rangle\langle\psi_i|=\mathds{1}_{N}$. Now, if Eq.~(\ref{PN-decomp}) is written in this $|\psi_i\rangle$ basis, then the right hand side is completely diagonal. This implies that the representation of $\rho$ on the left-hand-side must also be diagonal in this basis, that is, $|\psi_{i}\rangle$'s must necessarily be the eigenvectors of $\rho$ with $\rho=\sum_{i=1}^{N} \lambda_i|\psi_i\rangle\langle\psi_i|$. Here, we have denoted the corresponding eigenvalues as $\lambda_{i}$ and have assumed $\lambda_{1}\geq\lambda_{2}\geq...\geq\lambda_{N}$. The Eq.~(\ref{PN-decomp}) therefore takes the form:
\begin{equation}
 \rho = \sum_{i=1}^{N-1}(\lambda_i-\lambda_N)|\psi_i\rangle \langle \psi_i|+(N\lambda_N) \frac{\mathds{1_{N}}}{N}.
\end{equation}
As the weightages $s_{i}=(\lambda_i-\lambda_N)$ are non-negative, the above decomposition is necessarily unique. 
We note that Eq.~(\ref{PN-visibility}) expresses $P_N$ in terms of the eigenvalues of $\rho$. Using this, and after a straightforward calculation, we obtain an expression for $P_N$ solely in term of the weightage of the pure parts given as
\begin{align}
P_N &=\sqrt{\frac{\sum_{i=1}^{N-1}\sum_{j=i+1}^N(\lambda_i-\lambda_j)^2}{(N-1)(\sum_{i=1}^N\lambda_i)^2}}=\sqrt{\frac{N\sum_{i=1}^{N-1}s_i^2 -( \sum_{i=1}^{N-1}s_i)^2}{N-1}}\notag\\
&=\sqrt{(\sum_{i=1}^{N-1}s_i)^2-\frac{2N}{N-1}\sum_{i=1}^{N-1}\sum_{j=i+1}^{N-1}s_is_j} \leqslant \sum_{i=1}^{N-1}s_i. \label{PN-purity}
\end{align}
The above equation expresses the weightage of pure part interpretation of $P_{N}$. Just as in the 2D case, we find that $\rho$ can be generated by mixing together a completely mixed state and $N-1$ pure states in a particular proportion. However, the difference is that whereas $P_{2}=s_{1}$ in the 2D case, for the ND case, we find $P_{N}\leq \sum_{i=1}^{N-1}s_{i}$. In other words, the total weightage of pure parts puts an upper bound on the intrinsic degree of coherence. Moreover, the bound is tight as in any ND space, there exist states with only two non-zero eigenvalues. For such states,the bound is saturated, i.e, $P_{N}= \sum_{i=1}^{N-1}s_{i}$.  

\section{Quantifying the intrinsic degree of coherence $P_{\infty}$ of infinite-dimensional states}
In this section, we extend $P_{N}$ to the $N\to\infty$ limit to quantify the intrinsic degree of coherence $P_{\infty}$ of infinite-dimensional states. The procedure is not quite as straightforward as computing the $N\to\infty$ limit of Eq.~(\ref{P_N}) due to the following reasons: Firstly, from the expression for $P_{N}$, we note that in general, $\lim_{N\to\infty} P_{N}$ may not exist. This is because certain infinite-dimensional states can be non-normalizable, in which case $\mathrm{Tr}(\rho^2)$ can diverge \cite{merzbacher1998wiley}. Secondly, owing to the fact that $N$ can take only integer values, even if $\lim_{N\to\infty} P_{N}$ exists, the generalization implicitly assumes the existence of a discrete or countably-infinite basis in the infinite-dimensional vector space. While this assumption is manifestly valid for the infinite-dimensional spaces spanned by the discrete OAM and photon number bases, its validity is not evident for the infinite-dimensional space spanned by the uncountably-infinite or continuous variable position and momentum bases. Here, we present rigorous derivation of $P_\infty$ for infinite-dimensional states. We show that for any normalized infinite-dimensional state $\rho$ in the orbital angular momentum (OAM), photon number, position and momentum bases, the expression for $P_\infty$ is given by $P_{\infty}=\sqrt{\mathrm{Tr}(\rho^2)}$. 

\subsection{Orbital Angular Momentum and Angle Representations}
We denote the OAM eigenstates as $|l\rangle$, where $l=-\infty,...,-1,0,1,...,\infty$, and the angle eigenstates as $|\theta\rangle$, where $\theta\in[0,2\pi)$. Owing to the Fourier relationship between the OAM and angle observables \cite{peggbarnett1990pra}, the eigenstates are related as
\begin{subequations}\label{oam-ang-basisvecs}
\begin{align}\label{oam-eigvec}
 |l\rangle&=\frac{1}{\sqrt{2\pi}}\int_{0}^{2\pi} e^{+il\theta}|\theta\rangle\,\mathrm{d}\theta,\\\label{ang-eigvec}
 |\theta\rangle&=\frac{1}{\sqrt{2\pi}}\sum_{l=-\infty}^{+\infty} e^{-il\theta}|l\rangle.
\end{align}
\end{subequations}
We note that in contrast with finite-dimensional vectors, infinite-dimensional vectors may be non-normalizable. For instance, it is evident from Eq.~(\ref{ang-eigvec}) that the angle eigenstate $|\theta\rangle$ is non-normalizable. 

We now consider a state $\rho$ written in the OAM basis as
\begin{equation}\label{oam-improper-state}
 \rho=\sum_{l=-\infty}^{+\infty}\sum_{l'=-\infty}^{+\infty} c_{ll'}|l\rangle\langle l'|.
\end{equation}
We rewrite the state $\rho$ of Eq.~(\ref{oam-improper-state}) in the limiting form 
\begin{equation}\label{oam-proper-state}
 \rho=\lim_{D\to\infty} \sum_{l=-D}^{+D}\sum_{l'=-D}^{+D} c_{ll'}|l\rangle\langle l'|.
\end{equation}
In essence, the above relation views the infinite-dimensional state $\rho$ as the $D\to\infty$ limit of a $(2D+1)$-dimensional state residing in the finite state space spanned by the OAM eigenstates $|l\rangle$ for $l=-D,...,-1,0,1,...,D$, where $D$ is an arbitrarily-large but finite integer. We now use Eq.~(\ref{P_N}) to compute $P_{2D+1}$ and evaluate $P_{\infty}=\lim_{D\to\infty}P_{2D+1}$ which yields
\begin{equation}\label{Pinf-oam-proper}
 P_{\infty}=\lim_{D\to\infty} \sqrt{\frac{(2D+1)\,\sum_{l=-D}^{+D}\sum_{l'=-D}^{+D}|c_{ll'}|^2-1}{2D}}.
\end{equation}
Now let us assume that $\rho$ is normalized, that is $\mathrm{Tr}(\rho)=\sum_{l=-\infty}^{+\infty} c_{ll}=1$. This implies that $\sum_{l=-\infty}^{+\infty}\sum_{l'=-\infty}^{+\infty}|c_{ll'}|^2=\mathrm{Tr}(\rho^2)\leq 1$. Under this condition, Eq.~(\ref{Pinf-oam-proper}) evaluates to
\begin{equation}\label{Pinf-oam-physical}
 P_{\infty}=\sqrt{\sum_{l=-\infty}^{+\infty}\sum_{l'=-\infty}^{+\infty}|c_{ll'}|^2}=\sqrt{\mathrm{Tr}(\rho^2)}.
\end{equation}
The above equation can be used to evaluate $P_{\infty}$ of a normalized state $\rho$. However, when $\rho$ is non-normalizable, such as the angle eigenstate $\rho=|\theta\rangle\langle \theta|$ of Eq.~(\ref{ang-eigvec}), the quantity $\mathrm{Tr}(\rho^2)$ diverges. In such cases, Eq.~(\ref{Pinf-oam-physical}) cannot be used to compute $P_{\infty}$. 

We now use the basis invariance of $P_{\infty}$ to derive its expression in terms of the angle representation of $\rho$. Using Eq.~(\ref{oam-eigvec}) to substitute for $|l\rangle$ and $\langle l'|$ into Eq.~(\ref{oam-improper-state}), it follows that $\rho$ has the angle representation 
\begin{equation}\label{angle-improper-state}
 \rho=\int_{0}^{2\pi}\int_{0}^{2\pi} W(\theta,\theta')\,|\theta\rangle\langle \theta'|\,\mathrm{d}\theta\,\mathrm{d}\theta',\vspace{-2mm}
\end{equation}
where the continuous matrix elements $W(\theta,\theta')$ are related to the coefficients $c_{ll'}$ as
\begin{equation}\label{oam-angle-improper-wktheorem}
 W(\theta,\theta')=\frac{1}{2\pi}\sum_{l=-\infty}^{+\infty} \sum_{l'=-\infty}^{+\infty} c_{ll'}\,e^{+i(l\theta-l'\theta')}.
\end{equation}
In the context of light fields, $W(\theta,\theta')$ is the angular coherence function, which quantifies the correlation between the field amplitudes at angular positions $\theta$ and $\theta'$ \cite{jha2011pra,kulkarni2017natcomm}. Assuming that $\rho$ is normalized, we have $\mathrm{Tr}(\rho)=\int_{0}^{2\pi}W(\theta,\theta)\,\mathrm{d}\theta=1$. Substituting Eq.~(\ref{angle-improper-state}) in Eq.~(\ref{Pinf-oam-physical}), we obtain
\begin{equation}\label{Pinf-physical-angle}
 P_{\infty}=\sqrt{\mathrm{Tr}(\rho^2)}=\sqrt{\int_{0}^{2\pi}\int_{0}^{2\pi}|W(\theta,\theta')|^2\,\mathrm{d}\theta\,\mathrm{d}\theta'}.
\end{equation}
The equations (\ref{Pinf-oam-physical}) and (\ref{Pinf-physical-angle}) can be used to compute $P_{\infty}$ of any normalized infinite-dimensional state in the OAM and angle representations. 

\subsection{Photon number representation}

The photon number eigenstates $|n\rangle$, where $n=0,...,\infty$ span an orthonormal and complete basis in the infinite-dimensional Fock space. It is known that like OAM and angle, the photon number and optical phase are conjugate observables. However -- owing to the fact that unlike the OAM eigenvalues, the photon number eigenvalues can take only non-negative integer values -- the optical phase eigenstates in the infinite state space are not orthonormal, and therefore do not constitute a well-defined basis \cite{susskind1964ppf}. For our purposes it is sufficient to restrict our attention to the photon number basis, and compute $P_{\infty}$ in an identical manner as we did previously for states in the OAM basis. We first consider a general state expressed in the photon number basis as
\begin{equation}
 \rho=\sum_{n=0}^{\infty}\sum_{n'=0}^{\infty} a_{nn'}|n\rangle\langle n'|.
\end{equation}
We rewrite the above state in the limiting form
\begin{equation}
 \rho=\lim_{D\to\infty}\sum_{n=0}^{D}\sum_{n'=0}^{D} a_{nn'}|n\rangle\langle n'|,
\end{equation}
where $D$ is an arbitrarily-large but finite positive integer. We  then compute $P_{\infty}$ of $\rho$ by using Eq.~(\ref{P_N}) to compute $P_{D+1}$ of a $(D+1)$-dimensional state in the limit $D\to\infty$ as
\begin{equation}\label{Pinf-proper-photnumber}
 P_{\infty}=\lim_{D\to\infty} \sqrt{\frac{(D+1)\,\sum_{n=0}^{D}\sum_{n'=0}^{D}|a_{nn'}|^2-1}{D}}.
\end{equation}
We assume that $\mathrm{Tr}(\rho)=\sum_{n=0}^{\infty}a_{nn}=1$, which implies $\sum_{n=0}^{\infty}\sum_{n'=0}^{\infty}|a_{nn'}|^2=\mathrm{Tr}(\rho^2)\leq 1$. Under this condition, Eq.~(\ref{Pinf-proper-photnumber}) reduces to the form
\begin{equation}\label{Pinf-photnum-physical}
P_{\infty}=\sqrt{\sum_{n=0}^{\infty}\sum_{n'=0}^{\infty}|a_{nn'}|^2}=\sqrt{\mathrm{Tr}(\rho^2)}. 
\end{equation}
 
\subsection{Position and Momentum Representations}
We now consider infinite-dimensional states in the continuous-variable position and momentum representations. For conceptual clarity, we present our analysis for a one-dimensional configuration space which is labeled by the co-ordinate $x$. The corresponding canonical momentum space is labeled by the co-ordinate $p$. A general state $\rho$ in the position basis is written as
\begin{equation}\label{position-state-improper}
 \rho=\int_{-\infty}^{+\infty}\int_{-\infty}^{+\infty} G(x,x')\,|x\rangle\langle x'|\,\mathrm{d}x\,\,\mathrm{d}x'.
\end{equation}
Similarly, in the momentum basis $\rho$ is given by
\begin{equation}\label{momentum-state-improper}
 \hspace{-8mm}\rho=\int_{-\infty}^{+\infty}\int_{-\infty}^{+\infty} \Gamma(p,p')\,|p\rangle\langle p'|\,\,\mathrm{d}p\,\,\mathrm{d}p'.
\end{equation}
The continuous matrix elements $G(x,x')$ and $\Gamma(p,p')$ represent the cross-correlation functions in the position and momentum representations, respectively. 

 We recall that the expressions (\ref{Pinf-oam-physical}) and (\ref{Pinf-photnum-physical}) for $P_{\infty}$ of states in the OAM and photon number bases were derived by viewing the infinite-dimensional state as the infinite integer limit of a finite-dimensional state. As the dimensionality was constrained to take only integer values, the derivations implicitly depended on the fact that the OAM and photon number bases are discrete, and hence countably-infinite. However in the present case, both the position and the momentum bases are continuous, that is, uncountably-infinite. Nevertheless, we now show that this issue can be circumvented by constructing a physically indistinguishable finite-dimensional state space for position and momentum variables. Our construction extensively draws on techniques developed previously by Pegg and Barnett for constructing finite-dimensional state spaces for the OAM-angle \cite{peggbarnett1990pra} and photon number-optical phase \cite{peggbarnett1988epl,peggbarnett1989pra} pairs of observables.

\subsubsection{Construction of a finite-dimensional space}
We consider an arbitrarily-large but finite region $[-p_{\rm max},p_{\rm max}]$ in momentum space as depicted in Fig.~2. We sample $(2D+1)$ equally-spaced momentum values $p_{j}$ in this region, where $j=-D,...,0,...,D$, with $D$ also being arbitrarily large but finite. The spacing between consecutive values is $\Delta p=p_{\rm max}/D$, which is made arbitrarily close to zero. Using the $(2D+1)$ orthonormal eigenstates $|p_j\rangle$ corresponding to the momentum eigenvalues $p_{j}=j\Delta p$, we develop a consistent $(2D+1)$-dimensional state space for position and momentum. We will compute $P_{\infty}$ for $\rho$ by first computing $P_{2D+1}$ of a $(2D+1)$-dimensional state and then taking the limit of $D\to\infty$ and $p_{\rm max}\to \infty$, subject to the condition that $1/\Delta p=D/p_{\rm max}\to \infty$.

To this end, we note that a momentum operator $\hat{p}$ must be a generator of translations in position space. Therefore, a position state $|x\rangle$ must satisfy \cite{merzbacher1998wiley}
\begin{equation}\label{pos-shift}
 \exp\left(-i\hat{p}\eta/\hbar\right)|x\rangle=|x+\eta\rangle.
\end{equation}
If we define $|x_{0}\rangle$ as the state corresponding to the origin, then
\begin{equation}\label{position-shift}
|x\rangle=\exp\left(-i\hat{p}x/\hbar\right)|x_{0}\rangle.
\end{equation}
Now, similarly a position operator $\hat{x}$ must be a generator of translations in momentum space. This implies that 
\begin{equation}\label{momentum-shift}
 \exp\left(+ip_{k}\hat{x}/\hbar\right)|p_{j}\rangle=|p_{j+k}\rangle,
\end{equation}
where the translations are cyclic such that $\exp\left(ip_{1}\hat{x}/\hbar\right)|p_{D}\rangle=|p_{-D}\rangle$. We now use the orthonormal states $|p_{j}\rangle$ and equations (\ref{pos-shift}) and (\ref{momentum-shift}) to derive the form of the corresponding position eigenstates in the $(2D+1)$-dimensional state space.

Let us suppose that $|x_{0}\rangle$ takes the general form, 
\begin{equation}\label{x0-vector}
 |x_{0}\rangle=\sum_{j=-D}^{+D} c_{j}|p_{j}\rangle.
\end{equation}
Evaluating $\exp(+ip_{k}\hat{x}/\hbar)|x_{0}\rangle$ by using Eq.~(\ref{momentum-shift}), we get
\begin{equation}
 |x_{0}\rangle=\sum_{j=-D}^{+D} c_{j}|p_{j+k}\rangle.
\end{equation}
Now since the above equation is true for all $k$, the coefficients $c_{j}$ are necessarily independent of $j$, and upon normalization, they become $c_{j}=(1/\sqrt{2D+1})$. Using Eq.~(\ref{position-shift}), we then obtain
\begin{equation}\label{xvector}
 |x\rangle=\sum_{j=-D}^{+D}\frac{e^{-ip_{j}x/\hbar}}{\sqrt{2D+1}}|p_{j}\rangle.
\end{equation}
The inner product $ \langle x|x'\rangle$ can therefore be written as
\begin{align}\notag
 \langle x|x'\rangle&=\sum_{j=-D}^{+D}\sum_{k=-D}^{+D}\frac{e^{+i(p_{j}x-p_{k}x')/\hbar}}{(2D+1)}\langle p_{j}|p_{k}\rangle \\
 &=\frac{1}{(2D+1)}\frac{\sin\left[(2D+1)(x-x')\Delta p/2\hbar\right]}{\sin\left[(x-x')\Delta p/2\hbar\right]}.
\end{align}
This implies that $\langle x|x'\rangle=0$ only when $(x-x')=2\pi\hbar n/\{(2D+1)\Delta p\}$, where $n$ is a non-zero integer. This orthogonality condition allows us to select an orthonormal basis comprising the basis vectors $|x_{m}\rangle$ corresponding to the positions
\begin{equation}\label{xmvalues}
 x_{m}=\frac{2\pi m\hbar}{(2D+1)\Delta p}.\hspace{4mm} (m=-D,...,0,...,D)
\end{equation}
\begin{figure}[t!]
\centering
\includegraphics[width=85mm]{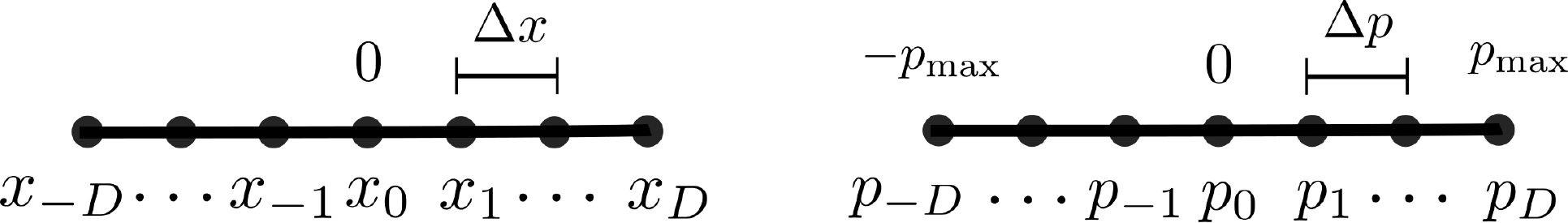}\label{pos-mom-fig}
\caption{In the finite state space, the position eigenvectors $|x_{m}\rangle$ for $m=-D,...,0,...D$, and momentum eigenvectors $|p_{j}\rangle$ for $j=-D,...,0,...,D$ span a finite $(2D+1)$-dimensional space.}
\end{figure}
These $(2D+1)$ positions are equally-spaced from $x_{-D}$ to $x_{D}$ with a spacing of $\Delta x=2\pi\hbar/\{(2D+1)\Delta p\}$. We write the orthonormality and completeness relations for the basis vectors $|x_{m}\rangle$ and $|p_{j}\rangle$ as 
\begin{subequations}\label{xp-orthocomp-proper}
\begin{align}
  &\langle x_{m}|x_{n}\rangle=\delta_{mn}, \hspace{14mm} \langle p_{j}|p_{k}\rangle=\delta_{jk},\\
  &\sum_{m=-D}^{+D} |x_{m}\rangle\langle x_{m}|=1,\hspace{10mm} \sum_{j=-D}^{+D} |p_{j}\rangle\langle p_{j}|=1.
 \end{align}
 \end{subequations}
Using equations (\ref{xvector}) and (\ref{xmvalues}), we find that the basis vectors are related as
 \begin{subequations}\label{xp-relations-proper}
 \begin{align}
 |x_{m}\rangle&=\frac{1}{\sqrt{2D+1}}\sum_{j=-D}^{+D} e^{-i2\pi mj/(2D+1)}\,|p_{j}\rangle,\\
 |p_{j}\rangle&=\frac{1}{\sqrt{2D+1}}\sum_{m=-D}^{+D} e^{+i2\pi mj/(2D+1)}\,|x_{m}\rangle.
 \end{align}
\end{subequations}
Thus, we have derived a finite-dimensional state space for position and momentum, which is depicted schematically in Fig.~1. In order to prove that the finite state space is physically consistent, we must show that the commutator $[\hat{x},\hat{p}]$ in this space is physically indistinguishable from the improper commutation relation $[\hat{x},\hat{p}]=i\hbar$. To this end, we note that $\hat{x}=\sum_{m=-D}^{+D}x_{m}|x_{m}\rangle\langle x_{m}|$ and $\hat{p}=\sum_{j=-D}^{+D}p_{j}|p_{j}\rangle\langle p_{j}|$. Using these expressions, we find that the commutator $[\hat{x},\hat{p}]$ has the following matrix elements:
 \begin{subequations}\label{xp-commutator-matrixelements}
 \begin{align}\label{xp-comm-xbasis}
 \langle x_{m}|[\hat{x},\hat{p}]|x_{n}\rangle&=\frac{2\pi\hbar(m-n)}{(2D+1)^2}\sum_{j=-D}^{+D}j\, e^{i2\pi(m-n)j/(2D+1)},\\\label{xp-comm-pbasis}
 \langle p_{j}|[\hat{x},\hat{p}]|p_{k}\rangle&=\frac{2\pi\hbar(k-j)}{(2D+1)^2}\sum_{m=-D}^{+D}m\, e^{-i2\pi(j-k)m/(2D+1)}.
 \end{align}
\end{subequations}
We notice that the diagonal elements $\langle x_{m}|[\hat{x},\hat{p}]|x_{m}\rangle$ and $\langle p_{j}|[\hat{x},\hat{p}]|p_{j}\rangle$ are all zero. As a result, the trace of $[\hat{x},\hat{p}]$ is zero, as expected for any commutator of finite-dimensional operators. We evaluate the above equations (\ref{xp-commutator-matrixelements}) in the limit $D\to\infty$ using Mathematica \cite{mathematica}, and simplify to obtain 
\begin{subequations}
\begin{align}
 [\hat{x},\hat{p}]&=\lim_{D\to\infty} i\hbar\Big[1-(2D+1)|x_{(D+\frac{1}{2})}\rangle\langle x_{(D+\frac{1}{2})}|\Big],\\
 [\hat{x},\hat{p}]&=\lim_{D\to\infty} i\hbar\Big[1-(2D+1)|p_{(D+\frac{1}{2})}\rangle\langle p_{(D+\frac{1}{2})}|\Big].
 \end{align}
\end{subequations}
We find that when the expectation value of $[\hat{x},\hat{p}]$ is evaluated for any physical state, the contributions from the second term in the above expressions asymptotically vanish. In this limit, we recover the usual commutator $[\hat{x},\hat{p}]=i\hbar$ for infinite-dimensional operators. Thus, we have constructed a consistent finite-dimensional state space for position and momentum. 

\subsubsection{Derivation of the expression for $P_{\infty}$}
We write the state $\rho$ from Eq.~(\ref{position-state-improper}) in the position basis of the finite-dimensional state space as
\begin{equation}\label{position-state-proper}
 \rho=\lim_{D\Delta x\to\infty}\lim_{\Delta x\to 0}\sum_{m=-D}^{+D}\sum_{n=-D}^{+D} \bar{G}_{x_{m}x_{n}}|x_{m}\rangle\langle x_{n}|.
\end{equation}
 Similarly, $\rho$ can be written in the momentum basis as
\begin{equation}\label{momentum-state-proper}
 \rho=\lim_{D\Delta p\to\infty}\lim_{\Delta p\to 0}\sum_{j=-D}^{+D}\sum_{k=-D}^{+D} \bar{\Gamma}_{p_{j}p_{k}}|p_{j}\rangle\langle p_{k}|.
\end{equation}
As $\rho$ is normalized, we have $\sum_{m=-D}^{+D}\bar{G}_{x_{m}x_{m}}=\sum_{j=-D}^{+D}\bar{\Gamma}_{p_{j},p_{j}}=1$. We can compute $P_{\infty}$ for $\rho$ by first computing $P_{2D+1}$ in terms of $\bar{G}_{x_{m}x_{n}}$ and $\bar{\Gamma}_{p_{j}p_{k}}$,  and then evaluating its limiting value as $D\to\infty$ and $p_{\rm max}\to\infty$, subject to the constraint $D/p_{\rm max}\to \infty$. These limits together ensure that $\Delta x\to 0$ and $\Delta p \to 0$, such that $D\Delta x\to\infty$ and $D\Delta p\to \infty$. Thus, we can compute $P_{\infty}$ in terms of $\bar{G}_{x_{m}x_{n}}$ as
\begin{equation}\label{P-pos-proper}
 P_{\infty}=\lim_{D\Delta x\to\infty}\lim_{\Delta x\to 0}\sqrt{\frac{2D+1}{2D}\Big[\sum_{m,n}|\bar{G}_{x_{m}x_{n}}|^2-\frac{1}{2D+1}\Big]}.
 \end{equation}
 Similarly in terms of $\bar{\Gamma}_{p_{j}p_{k}}$, we have
 \begin{equation}\label{P-mom-proper}
 P_{\infty}=\lim_{D\Delta p\to\infty}\lim_{\Delta p\to 0}\sqrt{\frac{2D+1}{2D}\Big[\sum_{j,k}|\bar{\Gamma}_{p_{j}p_{k}}|^2-\frac{1}{2D+1}\Big]}.
\end{equation}
In order to derive the form of $P_{\infty}$ in terms of $G(x,x')$ and $\Gamma(p,p')$, we must obtain the relation of these continuous functions to their discrete counterparts $\bar{G}_{x_{m}x_{n}}$ and $\bar{\Gamma}_{p_{j}p_{k}}$, respectively. Now if $\rho$ is a physical state, then $G(x,x')$ and $\Gamma(p,p')$ must be continuous integrable functions normalizable to unity. Thus, the relation of $G(x,x')$ to $\bar{G}_{x_{m}x_{m}}$, and that of $\Gamma(p,p')$ to $\bar{\Gamma}_{p_{j}p_{k}}$, must be such that $\sum_{m=-D}^{+D}\bar{G}_{x_{m}x_{m}}=\sum_{j=-D}^{+D}\bar{\Gamma}_{p_{j},p_{j}}=1$ should imply $\int_{-\infty}^{+\infty} G(x,x)\,\mathrm{d}x=\int_{-\infty}^{+\infty} \Gamma(p,p)\,\mathrm{d}p=1$. We now consider the relations 
\begin{subequations}\label{prop-improp}
 \begin{align}\label{prop-improp-x}
  G(x_{m},x_{n})&=\lim_{D\Delta x\to\infty}\lim_{\Delta x\to 0} \bar{G}_{x_{m}x_{n}}/\Delta x,\\\label{prop-improp-p}
  \Gamma(p_{j},p_{k})&=\lim_{D\Delta p\to\infty}\lim_{\Delta p\to 0}\bar{\Gamma}_{p_{j}p_{k}}/\Delta p.
 \end{align}
\end{subequations}
Substituting the above relations in $\sum_{m=-D}^{+D}\bar{G}_{x_{m}x_{m}}=\sum_{j=-D}^{+D}\bar{\Gamma}_{p_{j},p_{j}}=1$ yields $\lim_{D\Delta x\to\infty}\lim_{\Delta x\to 0} G(x_{m},x_{m})\Delta x$ and $\lim_{D\Delta p\to\infty}\lim_{\Delta p\to 0}\Gamma(p_{j},p_{j})\Delta p=1$. These summations are equivalent to the integral relations $\int_{-\infty}^{+\infty} G(x,x)\,\mathrm{d}x=\int_{-\infty}^{+\infty} \Gamma(p,p)\,\mathrm{d}p=1$, which implies that equations (\ref{prop-improp}) are correct. Upon substituting Eq.~(\ref{prop-improp-x}) in Eq.~(\ref{P-pos-proper}), and Eq.~(\ref{prop-improp-p}) in Eq.~(\ref{P-mom-proper}) and simplifying, we obtain
\begin{align}\notag
 P_{\infty}&=\lim_{D\Delta x\to\infty}\lim_{\Delta x\to 0} \sqrt{\sum_{m,n=-D}^{+D} |G(m\Delta x,n\Delta x)|^2\,\Delta x\,\Delta x},\\\notag
P_{\infty}&=\lim_{D\Delta p\to\infty}\lim_{\Delta p\to 0} \sqrt{\sum_{j,k=-D}^{+D} |\Gamma(j\Delta p,k\Delta p)|^2\,\Delta p\,\Delta p}.
\end{align}
The above equations can be expressed in integral form as \cite{riemann-integral} 
\begin{subequations}\label{P-improper}
\begin{align}\label{P-pos-improper}
  P_{\infty}&=\sqrt{\iint_{-\infty}^{+\infty}|G(x,x')|^2\,\mathrm{d}x\,\mathrm{d}x'}=\sqrt{\mathrm{Tr}(\rho^2)},\\\label{P-mom-improper}
  P_{\infty}&=\sqrt{\iint_{-\infty}^{+\infty} |\Gamma(p,p')|^2\,\mathrm{d}p\,\mathrm{d}p'}=\sqrt{\mathrm{Tr}(\rho^2)}.
\end{align}
\end{subequations}
Moreover, in terms of the Wigner function representation $W(x,p)=(1/(\pi\hbar))\int_{-\infty}^{+\infty}\langle x+y|\hat{\rho}|x-y\rangle e^{-2ipy/\hbar}\,\mathrm{d}y$ of $\rho$ \cite{wigner1932pr}, the measure $P_{\infty}$ can be expressed as 
\begin{equation}\label{P-wigfunc-improper}
 P_{\infty}=\sqrt{\mathrm{Tr}(\rho^2)}=\sqrt{2\pi\hbar\iint_{-\infty}^{+\infty}W^2(x,p)\,\mathrm{d}x\,\mathrm{d}p}.
\end{equation}
We note that the form of $P_{\infty}$ in Eq.~(\ref{P-pos-improper}) is identical to a measure known as the "overall degree of coherence" that was introduced and employed by Bastiaans for characterizing the spatial coherence of partially coherent fields in a complete manner \cite{bastiaans1983josa,bastiaans1984josaa}. Here, we have derived the measure for general classical and quantum states in the position and momentum representations from an entirely distinct perspective. 

\section{Conclusion and Discussion}

In the context of two-dimensional partially polarized electromagnetic fields, the basis-independent degree of polarization $P_2$ can be used to quantify the intrinsic degree of coherence of two-dimensional states. The measure $P_2$ has six known interpretations: (i) it is the Frobenius distance between the state and the identity matrix, (ii) it is the norm of the Bloch-vector representing the state, (iii) it is the distance to the center of mass in a configuration of point masses, (iv) it is the maximum of the degree of coherence, (v) it is the visibility in a polarization interference experiment, and (vi) it is equal to the weightage of the pure part of the state. By generalizing the first three interpretations, past studies had derived analogous expressions for the intrinsic degree of coherence $P_N$ of $N$-dimensional (ND) states. Here, we extended the concepts of visibility, degree of coherence, and weightage of pure part to ND states, and showed that $P_{2}$ generalizes to $P_{N}$ with respect to these interpretations as well. While other yet-to-be-discovered interpretations may still exist, we showed that $P_{N}$ has all the known interpretations of $P_{2}$, and can therefore be regarded as the intrinsic degree of coherence of $N$-dimensional states.  Finally, we extended the formulation of $P_{N}$ to the $N\to\infty$ limit and quantify the intrinsic degree of coherence $P_{\infty}$ of infinite-dimensional states in the OAM, photon number, position and momentum representations. 

\section*{Acknowledgment}
We thank Shaurya Aarav and Ishan Mata for discussions. We further acknowledge financial
support through grant no. EMR/2015/001931 from the Science and
Engineering Research Board, Department of Science \& Technology,
Government of India and through grant no. DST/ICPS/QuST/Theme -1/2019
from the Department of Science \& Technology, Government of India.

\bibliography{intcombo}

\begin{thebibliography}{53}
\bibitem [{rie()}]{riemann-integral}%
  \BibitemOpen
  \href@noop {} {}\bibinfo {note} {The Riemann integral is strictly defined
  only for bounded intervals. Here, the summation $\lim_{N\Delta x\to\infty}
  \lim_{\Delta x\to 0} \sum_{n=-N}^{+N} f(n\Delta x)\Delta x$ is first
  identified with the Cauchy principal value
  $\lim_{a\to\infty}\int_{-a}^{a}f(x)\,\mathrm{d}x$, which is then identified
  with the improper integral
  $\int_{-\infty}^{+\infty}f(x)\,\mathrm{d}x$.}\BibitemShut {Stop}%
\end{thebibliography}

\end{document}